\documentclass[12pt,a4paper,english]{article}

\usepackage[bindingoffset=0.3cm,textheight=22.5cm,hdivide={2.7cm,*,2.7cm}, vdivide={*,22cm,*}]{geometry}
\usepackage[bookmarksnumbered=true,breaklinks=true]{hyperref}

\usepackage{amsmath,amsfonts,amssymb,babel,slashed,graphicx,color}

\usepackage{mathtools,amssymb,bbm,slashed,graphicx,color,epstopdf,caption,subcaption,pdfpages,hyperref}

\newcommand{\kbar}{\mathchar'26\mkern-9mu k}

\usepackage[numbers,square,comma,sort&compress]{natbib}

\IfFileExists{dsfont.sty}
	{\usepackage{dsfont}
         \newcommand{\id}{\mathds{1}}}
	{\typeout{Package dsfont.sty was not found, using alternative macros.}
         \let\mathds=\mathbb
         \newcommand{\id}{\mbox{1 \kern-.59em {\rm l}}}}

\makeatletter

         \@addtoreset{equation}{section}
\makeatother

\newcommand{\nocontentsline}[3]{}
\newcommand{\tocless}[3]{\bgroup\let\addcontentsline=\nocontentsline#1{#2}#3\egroup}

\newcommand{\qed}{\nobreak \ifvmode \relax \else
      \ifdim\lastskip<1.5em \hskip-\lastskip
      \hskip1.5em plus0em minus0.5em \fi \nobreak
      \vrule height0.75em width0.5em depth0.25em\fi}

\newcommand{\be}{\begin{equation}}
\newcommand{\ee}{\end{equation}}
\newcommand{\eq}[1]{(\ref{#1})}

\def\nn{\nonumber}
\def\bea{\begin{eqnarray}}
\def\eea{\end{eqnarray}}

\def\beqa{\begin{eqnarray}} 
\def\eeqa{\end{eqnarray}} 
\def\beq{\begin{equation}} 
\def\eeq{\end{equation}}

 \def\L{\Lambda}

\def\s{\sigma}

 \def\cN{{\cal N}} 
 \def\cQ{{\cal Q}}

\newcommand{\R}{\mathds{R}}
\newcommand{\C}{\mathds{C}}
\newcommand{\N}{\mathds{N}}

\newcommand{\mmu}{\mathfrak{u}}

\def\bit{\begin{itemize}}
\def\eit{\end{itemize}}

\def\({\left(}
\def\){\right)}

\def\del{\partial}

 \allowdisplaybreaks[3]

\begin{document}

\renewcommand{\title}[1]{\vspace{10mm}\noindent{\Large{\bf
#1}}\vspace{8mm}} \newcommand{\authors}[1]{\noindent{\large
#1}\vspace{5mm}} \newcommand{\address}[1]{{\itshape #1\vspace{2mm}}}


\begin{flushright}
UWThPh-2015-06 
\end{flushright}

\begin{center}

\title{ \Large The squashed fuzzy sphere, fuzzy strings and the Landau problem}

\vskip 3mm

\authors{Stefan Andronache{\footnote{andronache.stefan@gmail.com}}, Harold C. Steinacker{\footnote{harold.steinacker@univie.ac.at}}
}
 
\vskip 3mm

 \address{ 

{\it Faculty of Physics, University of Vienna\\
Boltzmanngasse 5, A-1090 Vienna, Austria  }  
  }

\vskip 1.4cm

\textbf{Abstract}

\end{center}

We discuss the squashed fuzzy sphere, which is a projection of the fuzzy sphere onto the equatorial plane, 
and use it to illustrate the stringy aspects of noncommutative field theory.
We elaborate explicitly how strings linking its two coincident sheets arise in terms of fuzzy spherical harmonics.
In the large $N$ limit, the matrix-model Laplacian is shown to correctly reproduce the semi-classical dynamics 
of these charged strings, as given by the Landau problem.

\section{Introduction}\label{sec:background}

Field theory on noncommutative (NC) spaces has been studied intensively from various 
points of view in the past decades.
One of the original motivations was the (naive) hope that the UV-divergences of quantum field theory would be
regularized on a noncommutative space, due to the presence of 
an intrinsic noncommutative scale $\L_{NC}$. This hope turned out not to be vindicated. 
Rather, NC field theory behaves very differently 
from ordinary field theory at scales far above $\L_{NC}$, where the basic degrees of freedom
display a  string-like or dipole-like nature. 
This is already implicit in the trivial observation that NC fields are matrices or operators, 
which thus have two indices, and 
are naturally represented in t'Hoofts double line notation \cite{'tHooft:1974hx}.
Indeed, scalar fields on a noncommutative space arise in string theory as 
open strings starting and ending on a D-brane with $B$ field 
\cite{Chu:1998qz,Seiberg:1999vs}. This  suggests a 
dipole-like nature of noncommutative fields \cite{SheikhJabbari:1999vm,Bigatti:1999iz},
which is also implicit in the matrix-model realization of noncommutative gauge theory and 
its relation with string theory \cite{Alekseev:2000fd}, culminating
in the remarkable proposals \cite{Banks:1996vh,Ishibashi:1996xs} 
that string theory might be {\em defined} in terms of matrix models.
In particular, the IKKT matrix model is tantamount to noncommutative $\cN=4$ SYM on $\R^4_\theta$.

In the same vein, the  interactions determined by the algebra of noncommutative scalar fields 
with momentum far above $\L_{NC}$ is also very different from the commutative case; 
this can be seen easily on the quantum plane $\R^2_\theta$, but also e.g. on the fuzzy sphere
\cite{Chu2001}.
Since all these high-scale modes are probed in QFT via loop contributions, it should not be surprising 
that NC quantum field theory 
(NCQFT) is typically quite different from ordinary  QFT, and 
seems consistent only for very special models\footnote{This includes the
maximally supersymmetric $\cN=4$ Super-Yang-Mills,
which is nothing but the IKKT model, and a particular matrix model 
interpreted as scalar field theory \cite{Grosse:2012uv}.} .
The stringy nature of NCQFT manifests itself also in
the gravitational aspects of noncommutative gauge theory 
\cite{Rivelles:2002ez,Yang:2006hj,Steinacker:2007dq,Steinacker:2010rh} and the notorious UV/IR mixing \cite{Minwalla:1999px}.

These insights are very useful also to study NC field theory per se,
without any direct relation with string theory.
It allows to understand better its intrinsic properties, and 
 suggests a different organization of its fundamental degrees of freedom. 
In the present paper, we provide a 
particularly simple and explicit illustration of the stringy nature of noncommutative scalar fields,
in the example of noncommutative scalar
field theory on the squashed fuzzy sphere $P S^2_N$. This is a noncommutative space obtained by 
projecting the fuzzy sphere $S^2_N$ on the equatorial plane. It should be viewed as a stack of two coinciding
fuzzy disks with opposite (non-constant) Poisson structures,  glued together at their boundary.
The dipole or string picture discussed above suggests that there should be string-like modes connecting these 
two sheets, with opposite charges at the ends moving in the fields $B_+$ and $B_-$ on the two sheets.
Here $B_+ = - B_-$ corresponds to the symplectic forms i.e. the inverse Poisson structures on the two sheets.
At low energies, these should behave like point-like charged objects moving in an effective magnetic field
$B_+ - B_-$, which -- focusing on the center of the disks in a suitable scaling limit -- 
should reduce to the Landau problem\footnote{For a treatment of the Landau problem on the fuzzy 
sphere with monopole charge see \cite{Nair:2000ii,Karabali:2001te}. 
This is not directly related to the problem under investigation here.}.

With this in mind, we study  free scalar field theory on  $P S^2_N$, 
and identify the low-energy modes and their effective action. We can indeed identify the lowest eigenmodes
of the (matrix) Laplacian with string-like modes connecting the opposite sheets,
which reproduce precisely the energy levels and degeneracies of the Landau problem.
They are identified as fuzzy spherical harmonics $\hat Y^l_m$ with large quantum numbers $m \approx \pm l$.
For the lowest Landau level, the modes at (or near) the origin can be expressed 
succinctly in terms of coherent states
localized at the origin of the two sheets, thus exhibiting their stringy nature.
This is also related to recent results on the low-energy modes of coinciding or intersecting branes 
on squashed $SU(3)$ branes \cite{Steinacker:2014lma}.

The present paper hence demonstrates how 
an  appropriate organization of the degrees of freedom\footnote{A somewhat related organization of fields in terms of the
so-called the matrix base was used in \cite{Grosse:2004yu} to analyze
perturbation theory for scalar field theory on the quantum plane. } can illuminate the 
stringy physics hidden in NCFT, which  transcends the picture of conventional field theory.

\section{The Landau levels\label{sec:The-Landau-problem}}

We recall the quantum mechanical description of a (spinless) charged particle moving
perpendicular to an uniform magnetic field along  the $z$-axis,
\[
\vec{B}=B\hat{e}_{z} .
\]
The Hamiltonian for such a set-up is 
\begin{equation}
H=\frac{1}{2\mu}\left(\vec{P}-\frac{q}{c}\vec{A}\right)^{2}
\label{eq:Hamiltonian}
\end{equation}
where $\vec{A}$ is the vector potential related to the magnetic field, 
which has the  form

\begin{equation}
\vec{A}=\frac{B}{2}\begin{pmatrix}-Y\\
X\\
0
\end{pmatrix}\label{eq:gauge}
\end{equation}
in the Landau gauge.
Inserting (\ref{eq:gauge}) into (\ref{eq:Hamiltonian}) and introducing
the cyclotron (or Larmor) frequency
\begin{equation}
\omega_{c}=\frac{-qB}{\mu c}\label{eq:omega_c}
\end{equation}
the Hamiltonian can be written as
\[
H=\frac{P_{x}^{2}+P_{y}^{2}}{2\mu}+\frac{\mu\omega_{c}^{2}}{8}\left(X^{2}+Y^{2}\right)+\frac{\omega_{c}}{2}L_{z}=H_{xy}+\frac{\omega_{c}}{2}L_{z},
\]
where $L_{z}$ is the angular momentum operator in $z$-direction,
and $H_{xy}$ is the Hamiltonian of a two dimensional harmonic oscillator
with frequency $\frac{\omega_{c}}{2}$.
We can reformulate the problem in terms of the ladder operators
\begin{align*}
a_{r} & =\frac{1}{2}\left(\beta\left(X-iY\right)+\frac{i}{\beta\hbar}\left(P_{x}-iP_{y}\right)\right),\\
a_{l} & =\frac{1}{2}\left(\beta\left(X+iY\right)+\frac{i}{\beta\hbar}\left(P_{x}+iP_{y}\right)\right),
\end{align*}
with $\beta=\sqrt{\frac{\mu\omega_{c}}{2\hbar}}.$ These are the annihilation
operators of  right and left circular quanta respectively. We introduce
the number operators
\begin{eqnarray*}
N_{r} & = & a_{r}^{\dagger}a_{r},\\
N_{l} & = & a_{l}^{\dagger}a_{l}
\end{eqnarray*}
so that
\begin{eqnarray*}
H_{xy} & = & \left(N_{r}+N_{l}+1\right)\frac{\hbar\omega_{c}}{2},\\
L_{z} & = & \left(N_{r}-N_{l}\right)\hbar.
\end{eqnarray*}
 Now it is evident that $a_{r}^{\dagger}$ ($a_{l}^{\dagger}$)
create right (left) circular quanta. Both raise the energy by $\frac{\hbar\omega_{c}}{2}$,
but acting with $a_{r}^{\dagger}$ increases the additional
angular momentum by  $\hbar$, while acting with $a_{l}^{\dagger}$ decreases the
angular momentum by $\hbar$.
Thus the Hamiltonian (\ref{eq:Hamiltonian}) has the form
\begin{align*}
H & =\left(N_{r}+\frac{1}{2}\right)\hbar\omega_{c}
\end{align*}
with eigenvalues 
\begin{equation}
E=\left(n_{r}+\frac{1}{2}\right)\hbar\omega_{c}\label{eq:E_perp n_r}
\end{equation}
and eigenfunctions

\[
\left|\chi_{n_{r},n_{l}}\right\rangle =\frac{1}{\sqrt{n_{r}!n_{l}!}}(a_{r}^{\dagger})^{n_{r}}(a_{l}^{\dagger})^{n_{l}}\left|\chi_{0,0}\right\rangle,
\qquad n_l, n_r \in \N.
\]
Note that the energy depends only on $n_r$ but is independent
of $n_{l}$, thus the energy states
corresponding to a particular Landau level  $n_{r}$ are infinitely degenerate.
It is not hard to see that the wave-functions for the lowest Landau level $n_r=0$
are concentric circles around the origin with radius measured by $n_l$,
\begin{align}
\chi_{0,n_l}(\rho,\varphi)=\frac{\beta}{\sqrt{\pi n_l!}} e^{-in_l \varphi} (\beta \rho)^{n_l} e^{\frac{-\beta^{2}\rho^{2}}{2}}
\label{landau-wavefunct}
\end{align}
in polar coordinates.

If we include spin, the Hamiltonian is modified as follows
\begin{equation}
H=\left(N_{r}+\frac{1}{2}-\frac{\sigma_{z}}{2}g\right)\hbar\omega_{c}\label{eq:landauwithspin}
\end{equation}
where $\sigma_{z}$ is the spin operator in $z$-direction and $g$
the g-factor dependent on the type of particle.

\section{The fuzzy sphere $S_{{N}}^{2}$}
\label{part:The-fuzzy-sphere}

The fuzzy sphere $S_{{N}}^{2}$ \cite{madore,hoppe}
is a quantization of the usual sphere $S^{2}$
with a cutoff in angular momentum, which contains $N$ quanta of area.
The quantization of $S^{2}$ is given by a quantization map $\mathcal{Q}$,
\begin{eqnarray}
\mathcal{Q}: & \quad\mathcal{C}_{n}(S^{2}) & \rightarrow\mathcal{M}_{_{N}}=\mathrm{Mat}(N,\mathbb{C})\label{eq:x->X}\\
 & x^{a} & \mapsto\; X^{a}=\kappa J^{a}\nonumber 
 \label{X-embedding-J}
\end{eqnarray}
which maps in particular the embedding functions $x^{a}$ on $S^{2}$ to quantized
embedding functions $X^{a}=\kappa J^{a}$ on $S_{N}^{2}$. 
Here $J^{a}$ are the generators of $\mathfrak{su}(2)$ in the 
$N=2n+1$-dimensional irreducible representation, $\mathcal{C}_{n}(S^{2})$
is the space of polynomials on $S^2$ of degree $\leq n$ and
$\mathcal{M}_{_{N}}$ is the algebra of complex $N\times N$ matrices.
Since the quadratic Casimir operator has the form
\[
\vec{J}^{2}=C_{N}\mathbbm{1}\qquad\text{{with}}\quad C_{N}=\frac{1}{4}\left(N^{2}-1\right),
\]
the radial constraint of a sphere with radius  $r$
\[
\left(X^{1}\right)^{2}+\left(X^{2}\right)^{2}+\left(X^{3}\right)^{2}=r^{2}
\]
is recovered if we set
\[
\kappa^{2}=\frac{r^{2}}{C_{N}} .
\]
We  introduce a constant which is the analogue of $\hbar$
\begin{align}
 \kbar=\kappa r=\frac{r^{2}}{\sqrt{C_{N}}}
 \label{kbar-def}
\end{align}
and the commutative limit is given by $\kbar\rightarrow0$ as $N\rightarrow\infty$
for fixed radius. The generators $X^{a}$ of the algebra $\mathcal{M}_{_{N}}$
satisfy the commutation relations
\begin{eqnarray}
\left[X^{a},X^{b}\right] & = & i\kbar C_{\;\;\; c}^{ab}X^{c}=:i\Theta^{ab},\\
C^{abc} & = & r^{-1}\varepsilon^{abc},\\
(\Theta^{ab})_{S_{N}^{2}} & = & \frac{\kbar}{r}\begin{pmatrix}0 & X^{3} & -X^{2}\\
-X^{3} & 0 & X^{1}\\
X^{2} & -X^{1} & 0
\end{pmatrix}.
\label{commrel-S2}
\end{eqnarray}
To complete the definition of the quantization map $\cQ$, 
we  decompose $\mathcal{M}_{_{N}}$ into irreducible representations
under the adjoint action of $\mathfrak{su}(2)$ 
\begin{eqnarray}
\mathcal{M}_{_{N}}\cong(N)\otimes(\bar{N}) & = & (1)\oplus(3)\oplus\ldots\oplus(2N-1)\nonumber \\
 & = & \left\{ \hat{Y}_{0}^{0}\right\} \oplus\ldots\oplus\left\{ \hat{Y}_{m}^{N-1}\right\} .\label{eq:Mn decomp}
\end{eqnarray}
This defines the  fuzzy spherical harmonics $\hat{Y}_{m}^{l}$, and 
allows to write down a natural definition for the quantization
map $\mathcal{Q}$ for polynomial functions of degree less than or equal to $n=2N+1$:
\begin{eqnarray*}
\mathcal{Q}:\quad\mathcal{C}_{n}(S^{2}) & \rightarrow & \mathcal{M}_{_{N}}=\mathrm{Mat}(N,\mathbb{C})\\
Y_{m}^{l} & \mapsto & \hat{Y}_{m}^{l} ,
\end{eqnarray*}
compatible with the $SO(3)$ symmetry. Here $Y_{m}^{l}$ are the usual spherical harmonics.
In the limit $N \to \infty$, we recover the full algebra of polynomial functions on $S^2$.

The commutation relations \eq{commrel-S2} define a quantization of the Poisson structure 
\begin{eqnarray*}
\{x^{a},x^{b}\} & = & \kbar C_{\;\;\; c}^{ab}x^{c}=:\theta^{ab},\\
C^{abc} & = & r^{-1}\varepsilon^{abc} \\
\theta^{ab} & = & \frac{\kbar}{r}\begin{pmatrix}0 & x^{3} & -x^{2}\\
-x^{3} & 0 & x^{1}\\
x^{2} & -x^{1} & 0
\end{pmatrix}
\end{eqnarray*}
which corresponds to the  $SO(3)$--invariant  symplectic 2-form
\begin{equation}
\omega_N=\frac{1}{\kbar}C_{abc}x^{a}\mathrm{d}x^{b}\wedge\mathrm{d}x^{c}
\label{eq:omega sympl}
\end{equation}
and satisfies the flux quantization condition $2\pi N=\intop_{S^{2}}\omega_N$.
Thus the fuzzy sphere $S^2_N$ is the quantization of the symplectic manifold $(S^{2},\omega_N)$.
Furthermore, the Laplace operator on the fuzzy sphere is  defined by 
\begin{equation}
\Box=\frac{1}{\kbar^{2}}\sum_{a=1}^3\left[X^{a},\left[X^{a},.\right]\right].
\label{eq:BOX-sphere}
\end{equation}
This type of matrix Laplacian arises naturally in the context of Yang-Mills 
models\footnote{For example in the  IKKT model \cite{Ishibashi:1996xs}, 
the matrices $X^a$ transform in the adjoint of some $U(N)$ gauge group.
Assuming that they acquire non-trivial expectation values such as $X^a \sim J^a$ \eq{X-embedding-J},
the $U(N)$ gauge symmetry is spontaneously broken, and linearized transversal fluctuations on such a background are 
governed by the Laplace operator \eq{eq:BOX-sphere}.
This can be viewed as a variant of the Higgs mechanism. Here, we simply take \eq{eq:BOX-sphere}
 as a natural starting point, ignoring possible extra degrees of freedom which may arise in other contexts.
} 
\cite{Steinacker:2010rh}.

\subsection{Fuzzy spherical harmonics\label{sub:fuzzy-spherical harmonics}}

The fuzzy spherical harmonics $\hat{Y}_{m}^{l}$ were identified 
in equation (\ref{eq:Mn decomp}) as the irreducible representations
of $SU(2)$ acting on the non-commutative algebra $\mathcal{M}_{_{N}}$, analogous
to the commutative case up to a cutoff. It is easy to see that they are also eigenfunctions
of the Laplace operator
\begin{equation}
\Box\hat{Y}_{m}^{l}=\frac{\kappa^{2}}{\kbar^{2}}l(l+1)\hat{Y}_{m}^{l}=\frac{1}{r^{2}}l(l+1)\hat{Y}_{m}^{l} ,
\label{eq:Box EV}
\end{equation}
in analogy to the classical case,  with the same $2l+1$-fold degeneracy.
We can get more  information on the explicit (matrix) form of the $\hat{Y}_{m}^{l}$
for fixed $N$ using the representation theory of $SU(2)$. 
Consider a basis where the Cartan generator $H$ of $SU(2)$
is diagonal. Since $m$ gives the eigenvalue of $H$, 
all  the matrices $\hat{Y}_{0}^{l}$ are diagonal,  $\hat{Y}_{1}^{l}$
have entries only along the first diagonal above the main diagonal,
$\hat{Y}_{2}^{l}$ have entries only along the second diagonal above
the main diagonal and so forth. An analogous statement can be made
for $\hat{Y}_{-1}^{l}$, $\hat{Y}_{-2}^{l}$, etc. below the main
diagonal. The entries of the matrices are symmetric w.r.t. the anti-diagonal, and
their values are decreasing with increasing distance from
the anti-diagonal. Clearly the maximal value for $l$ is
$l_{max}=N-1$, and all matrices with $\left|m\right|>l$ vanish.

\section{The squashed fuzzy sphere $PS^2_N$\label{sec:The-fuzzy-disk}}

In this section we discuss the squashed fuzzy sphere, which is interpreted 
as projection of the fuzzy sphere onto the equatorial plane \cite{Steinacker:2014lma}. 
This arises e.g. as building block of cosmological solutions in the IR-regulated 
IKKT matrix model \cite{Kim:2012mw}.
In particular, we explain
how strings linking its two coincident sheets arise in terms of noncommutative functions.
The relation of matrix models with noncommutative gauge theory is illustrated by showing 
how the description of these strings in noncommutative field theory reproduces the
semi-classical dynamics of these charged strings as given by the Landau problem.

A projection $\Pi$ of a classical sphere onto its equatorial plane\footnote{Note that we are considering 
an orthogonal projection rather than a stereographic projection here.} 
is achieved simply by 
replacing the three embedding functions $x^a: \ S^2 \hookrightarrow \R^3$
by only two embedding functions $x^1$ and $x^2$, dropping $x^3$:
\begin{equation}
\begin{array}{rcl}
S^2    &\rightarrow  \quad \R^3  \quad  &  \stackrel{\Pi}{\rightarrow}\quad \R^2  \\
p  \   &\mapsto \quad  x^a(p)\quad & \mapsto \quad x^a(p),\quad a=1,2 
\end{array}
\end{equation}
Here we keep the same space of functions on $S^2$, 
but change the embedding information given by the $x^a$.
After projecting, the
two hemispheres are stacked one onto another as two coinciding disks glued at the boundary.

Accordingly, we define the projected or squashed fuzzy sphere  $PS^2_N$
in terms of the {\em two} generators  $X^a,\ a=1,2$. They generate the same algebra
of fuzzy functions $\mathrm{Mat}(N,\mathbb{C})$ as for $S^2_N$,
but will lead to a different fuzzy Laplacian.
It can be viewed as two projected fuzzy disks glued at the boundary. 
The relation between the fuzzy disk and the fuzzy sphere can be seen 
explicitly by expressing $X^3$ in terms of the two independent generators $X^1, X^2$:
\begin{equation}
(X^{1})^{2}+(X^{2})^{2}+(X^{3})^{2}=r^{2}\quad\Rightarrow\quad
(X^{3})^2=r^{2}-(X^{1})^{2}-(X^{2})^{2}.\label{eq:elimination of X_3}
\end{equation}
\begin{figure}[h]
\centering{}\includegraphics[scale=0.13]{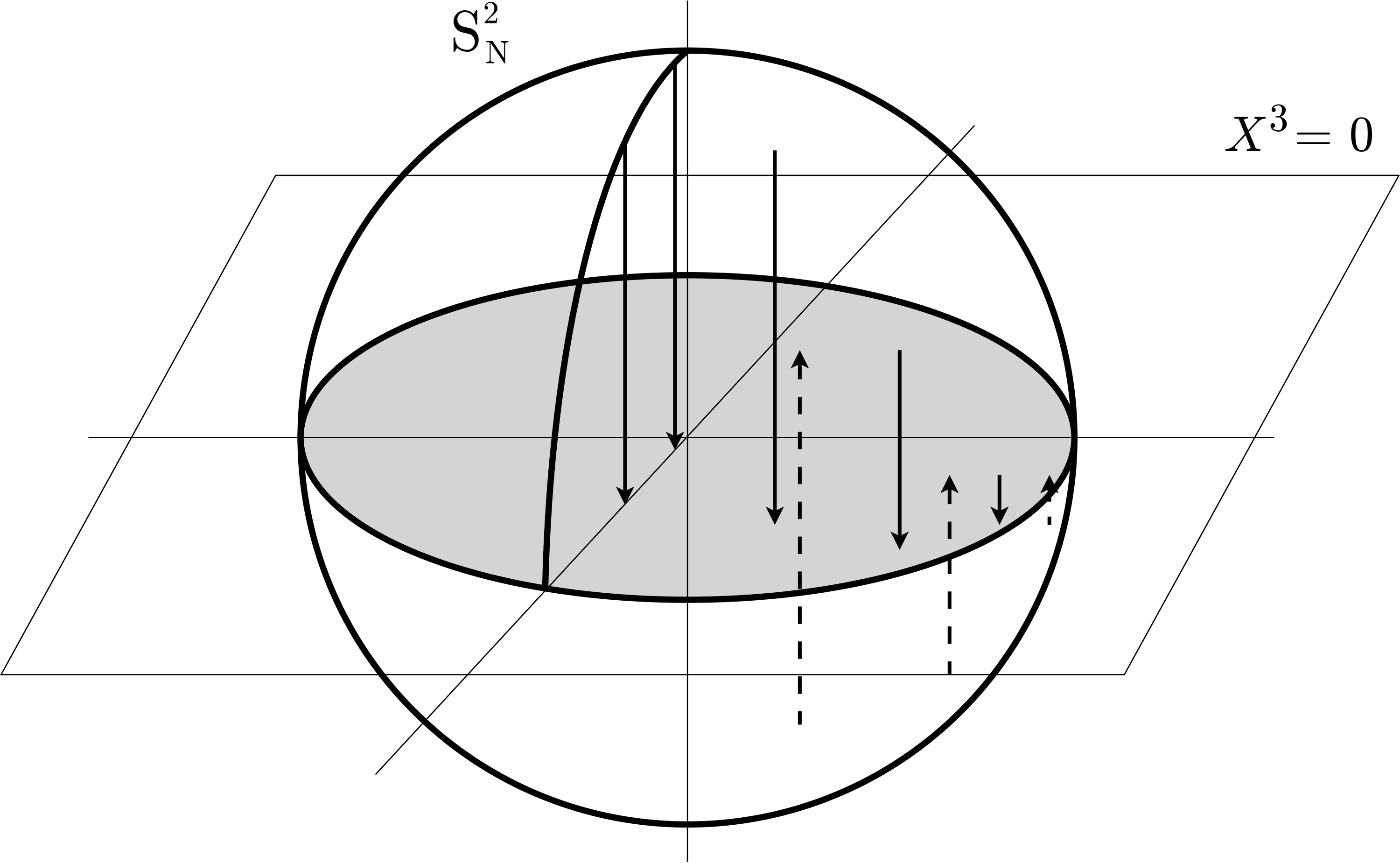}
\caption{A schematic depiction of the orthographic projection $\Pi$ of the fuzzy
sphere onto the $X^{3}=0$ plane. The solid arrows indicate projections
of the upper hemisphere, while the dashed arrows are projections from
the lower hemisphere. The shaded area is the area onto which is projected,
i.e. the interleaved fuzzy disks.}
\end{figure}

We define 
\begin{equation}
 X^3_\pm = \pm\sqrt{r^{2}-(X^{1})^{2}-(X^{2})^{2}}
\end{equation}
as positive respectively negative part of $X^3$.
Then $X^3_\pm$ reduces in the semi-classical (i.e. Poisson) limit to the embedding functions 
$x^3_\pm$ of upper respectively lower hemi-sphere in $\R^3$.
Then the matrix Laplacian on the squashed fuzzy sphere is 
\begin{equation}
\Box_S=\frac{1}{\kbar^{2}}\sum_{i=1}^2\left[X^{i},\left[X^{i},.\right]\right].
\label{eq:BOX-sqashed}
\end{equation}

\subsection{Poisson structure}

The commutators of the generators $X^1, X^2$ of the
squashed fuzzy sphere define in the semi-classical limit a Poisson structure on the 
projected disks. This is nothing but the push-forward of the Poisson structure on $S^2$ by $\Pi$. 
On the upper sheet, we have
\begin{equation}
 \{x^1,x^2\} = \frac{\kbar}{r} x^3_+(x^1,x^2) = \frac{\kbar}{r} \sqrt{r^{2}-(x^{1})^{2}-(x^{2})^{2}}
 = \theta_{+}^{12}
\end{equation}
while on the lower sheet we have 
\begin{equation}
 \{x^1,x^2\} = \frac{\kbar}{r} x^3_-(x^1,x^2) = -\frac{\kbar}{r} \sqrt{r^{2}-(x^{1})^{2}-(x^{2})^{2}} .
 = \theta_{-}^{12}
 \end{equation}
Thus the Poisson tensor on the two sheets indicated by $\pm$ is given by 
\begin{equation}
\theta_{\pm}^{ij}=\pm\frac{\sqrt{r^{2}-(x^{1})^{2}-(x^{2})^{2}}}{r}\kbar\left(\begin{array}{cc}
0 & 1\\
-1 & 0
\end{array}\right).\label{eq:theta disk}
\end{equation}
We observe that the two coinciding fuzzy disks have opposite Poisson structure 
\begin{equation}
-\theta_{+}^{ij}=\theta_{-}^{ij} ,
\label{eq:poisson-struture-pm}
\end{equation}
and  $\theta_\pm$ vanishes as we approach the edge, so that we have a smooth
transition 
\begin{equation}
r^{2}-(x^{1})^{2}-(x^{2})^{2}=0\quad\Rightarrow\quad\theta_{+}^{ij}=\theta_{-}^{ij}=0.
\label{eq:theta-edge}
\end{equation}
Of course the semi-classical treatment at the edge may be questioned, however this will not be 
important below.
We observe that both the Poisson structure and the Laplacian are quite different from the 
corresponding structures on the
single fuzzy disks defined in \cite{Lizzi:2003ru,Lizzi:2005zx} via a truncation of the quantum plane.

\subsection{Effective gauge fields}

In this section, we will  obtain an interpretation of the matrix Laplacian 
$\Box$ in terms of noncommutative gauge theory. This will allow to identify particular 
functions on the squashed fuzzy sphere as charged strings linking its two sheets, and provide an
explicit relation with the energy levels of Landau problem.

To understand this relation, we recall that gauge fields on the Moyal-Weyl quantum plane  $\R^2_\theta$
can be introduced as deformations 
\begin{equation}
X^{i}=\bar{X}^{i}-\bar{\theta}^{ij}A_{j}(\bar{X}) 
\label{cov-coords}
\end{equation}
 of the generators $\bar{X}^i$ of  $\R^2_\theta$, which satisfy
\begin{equation}
 [\bar{X}^{i},\bar{X}^{j}] = i \bar \theta^{ij} = i \kbar \begin{pmatrix}
                                                                0 & 1 \\ -1 & 0
                                                               \end{pmatrix} .
\end{equation}
The $X^i$ are known as covariant coordinates \cite{Madore:2000en}. 
Their commutators are given by
\begin{eqnarray} 
  [{X}^{i},{X}^{j}] & = & [\bar{X}^{i},\bar{X}^{j}] -\bar{\theta}^{jj'}[\bar{X}^{i},A_{j'}]
  +\bar{\theta}^{ii'}[ \bar{X}^{j},A_{i'}] 
  +\bar{\theta}^{ii'}\bar{\theta}^{jj'}[ A_{i'},A_{j'}] \nonumber \\
 & = & -i\bar{\theta}^{ii'}\bar{\theta}^{jj'}\left(\bar{\theta}_{i'j'}^{-1}-\partial_{i'}A_{j'}
  +\partial_{j'}A_{i'}+ i[ A_{i'},A_{j'}] \right)\nonumber \\
 & = &  i \bar \theta^{ij} -i\bar{\theta}^{ii'}\bar{\theta}^{jj'}F_{i'j'}
\end{eqnarray}
where $F_{ij}$ can be interpreted as field strength of the $U(1)$ gauge field\footnote{Recall that 
in noncommutative field theory, the field strength  contains commutators even for abelian 
i.e. $U(1)$ gauge fields. However these terms are subleading in the semi-classical limit, and will be dropped here.} 
$A_i$ on 
 $\R^2_\theta$. The commutators 
\begin{align}
 [X^i,\phi] &= [\bar X^i,\phi] -  {\bar\theta}^{ii'}[A_{i'},\phi] \nn\\
  &=  i \bar\theta^{ii'}(\del_{i'} + i [A_{i'},\phi]) \nn\\
   &= i \bar\theta^{ii'}D_{i'}\phi
     \label{covar-der}
\end{align}
defines the covariant derivatives of a scalar field $\phi$. 
 In the semi-classical limit, the Poisson-brackets of $x^i \sim X^i$ can be expressed accordingly 
\begin{eqnarray}
\left\{ x^{i},x^{j}\right\}  & = & \left\{ \bar{x}^{i},\bar{x}^{j}\right\} -\bar{\theta}^{jj'}\left\{ \bar{x}^{i},A_{j'}\right\} +\bar{\theta}^{ii'}\left\{ \bar{x}^{j},A_{i'}\right\} +\bar{\theta}^{ii'}\bar{\theta}^{jj'}\left\{ A_{i'},A_{j'}\right\} \nonumber \\
 & = &  \bar \theta^{ij} - \bar{\theta}^{ii'}\bar{\theta}^{jj'} F_{i'j'}\nonumber \\
 & {=} & \theta^{ij}
 \label{eq:{x_i,x_j}} 
\end{eqnarray}
as deformation of the constant Poisson bracket 
$\left\{ \bar{x}^{i},\bar{x}^{j}\right\} = \bar\theta^{ij}$ by
the field strength $F_{ij}$  
\[
F_{ij}=\partial_{i}A_{j}-\partial_{j}A_{i}-\left\{ A_{i},A_{j}\right\} .
\]
Thus $\bar x^i$ can be viewed as Darboux coordinates on $(\R^2,\{.,.\})$.
The semi-classical version of  \eq{cov-coords} 
\begin{equation}
x^{i}=\bar{x}^{i}-\bar{\theta}^{ij}A_{j}(\bar{x}) 
\end{equation}
therefore allows to interpret the difference between the $x^i$ and the Darboux coordinates 
$\bar{x}^{i}$ in terms of a $U(1)$ gauge field.

We now apply these insights to the example of the squashed fuzzy sphere. 
Since its finite-dimensional setting cannot strictly be viewed as a deformation 
of the quantum plane $\R^2_\theta$, 
we restrict ourselves to the semi-classical (i.e. Poisson) limit. 
More precisely, we consider the limit 
corresponding to $N\to\infty$, keeping the leading order in the noncommutativity scale $\kbar$. 
Higher powers in $\kbar$ can be neglected as long as 
the physical momenta are sufficiently low.

As we have seen in (\ref{eq:poisson-struture-pm}), the squashed
fuzzy sphere decomposes into an upper and a lower fuzzy disk, 
which arise by restricting the matrices to the upper and 
lower blocks defined by the positive and negative spectrum of $X^3$:
\begin{equation}
X^i=\begin{pmatrix}X_{+}^{i} & 0\\
0 & X_{-}^{i}
\end{pmatrix}
\sim \begin{pmatrix}x_{+}^{i} & 0\\
0 & x_{-}^{i}
\end{pmatrix}
=\begin{pmatrix}\bar{x}^{i}-\bar{\theta}^{ij}A_{j}^{+} & 0\\
0 & \bar{x}^{a}-\bar{\theta}^{ij}A_{j}^{-}
\end{pmatrix}
\label{eq:full X}
\end{equation}
with $+$ ($-$) indicating the upper (lower) sheet.
Note that although the full Poisson structures
$\theta^{ij}_+, \theta^{ij}_-$ have opposite sign, 
the Darboux coordinates $\bar{x}^{i}$ define the {\em same} constant $\bar\theta^{ij}$
on the upper and the lower sheet,
\begin{equation}
\left\{ \bar{x}^{i},\bar{x}^{j}\right\} =\bar{\theta}^{ij}=\kbar\varepsilon^{ij}.
\label{eq:Darboux bracket}
\end{equation}
This is essential for an interpretation in terms of 
noncommutative gauge theory on a stack of coinciding branes.

Now we want to find the corresponding gauge fields $A_i^\pm$
on the two sheets explicitly.
From (\ref{eq:theta disk}) we get 
\[
\left\{ x_{\pm}^{i},x_{\pm}^{j}\right\} =\theta_{\pm}^{ij}=\pm\frac{\sqrt{r^{2}-(x^{1})^{2}-(x^{2})^{2}}}{r}\kbar\varepsilon^{ij}
\]
which indeed reduces to (\ref{eq:Darboux bracket})
for $x^{1},x^{2}\simeq\vec{0}$, and vanishes
at the edge of the disk. We can rewrite equation (\ref{eq:{x_i,x_j}})
as 
\[
\bar{\theta}_{i'i}^{-1}\bar{\theta}_{j'j}^{-1}\theta^{ij}=-\bar{\theta}_{i'j'}^{-1}-F_{i'j'}
\]
with 
\[
\bar{\theta}_{ij}^{-1}=\frac{1}{\kbar}\left(\begin{array}{cc}
0 & -1\\
1 & 0
\end{array}\right)
\]
and obtain
\[
\bar{\theta}_{ii'}^{-1}\bar{\theta}_{jj'}^{-1}\theta^{i'j'}=\frac{1}{\kbar r}\left(\begin{array}{cc}
0 & \sqrt{r^{2}-\left(x^{1}\right)^{2}-\left(x^{2}\right)^{2}}\\
-\sqrt{r^{2}-\left(x^{1}\right)^{2}-\left(x^{2}\right)^{2}} & 0
\end{array}\right).
\]
We thus obtain the field strength $F$  on the different sheets as 
\begin{eqnarray*}
F & = & \begin{pmatrix}F^{+} & 0\\
0 & F^{-}
\end{pmatrix}
\end{eqnarray*}
with 
\[
F^{\pm}_{ij}
=\frac{1}{\kbar}\left(\begin{array}{cc}
0 & 1\mp\frac{1}{r}\sqrt{r^{2}-(x^{1})^{2}-(x^{2})^{2}}\\
-\left(1\text{\ensuremath{\mp}}\frac{1}{r}\sqrt{r^{2}-(x^{1})^{2}-(x^{2})^{2}}\right) & 0
\end{array}\right).
\] 
To obtain explicit expressions for the gauge fields $\vec{A}^{\pm}$, 
we have to solve the following differential equations 
\begin{equation}
F_{12}^{{\color{red}{\normalcolor \pm}}}=\kbar^{-1}\left(1\text{\ensuremath{\mp}}\frac{1}{r}\sqrt{r^{2}-(x^{1})^{2}-(x^{2})^{2}}\right)=\partial_{1}A_{2}^{\pm}-\partial_{2}A_{1}^{\pm}-\left\{ A_{1}^{\pm},A_{2}^{\pm}\right\} .\label{eq:F pm full}
\end{equation}
Since $\left\{ A_{1}^{\pm},A_{2}^{\pm}\right\} $ is of higher order
in $\kbar$ than $\partial_{1}A_{2}^{\pm}-\partial_{2}A_{1}^{\pm}$,  it is negligible
in the semi-classical limit, and (\ref{eq:F pm full})
simplifies to 
\begin{equation}
F_{12}^{\pm}\simeq\partial_{1}A_{2}^{\pm}-\partial_{2}A_{1}^{\pm}.\label{eq:F pm ohne poisson}
\end{equation}
The solutions of this differential equations are given by 
\begin{equation}
\vec{A}^{\pm}(\vec{x})=\frac{1}{2\kbar}\begin{pmatrix}-x^{2}\pm K^{1}\\
x^{1}\pm K^{2}
\end{pmatrix}\label{eq:A_pm}
\end{equation}
where 
\begin{eqnarray*}
K^{1} & = & \frac{1}{2r}\left(x^{2}\sqrt{r^{2}-(x^{1})^{2}-(x^{2})^{2}}+\left(r^{2}-(x^{1})^{2}\right)\arctan\left(\frac{x^{2}}{\sqrt{r^{2}-(x^{1})^{2}-(x^{2})^{2}}}\right)\right),\\
K^{2} & = & \frac{-1}{2r}\left(x^{1}\sqrt{r^{2}-(x^{1})^{2}-(x^{2})^{2}}+\left(r^{2}-(x^{2})^{2}\right)\arctan\left(\frac{x^{1}}{\sqrt{r^{2}-(x^{1})^{2}-(x^{2})^{2}}}\right)\right).
\end{eqnarray*}
It is  easy to verify 
\[
F_{12}^{\pm}=\vec{\nabla}\times\vec{A}^{\pm}.
\]
Now consider in more detail the covariant derivative \eq{covar-der} acting on general noncommutative 
scalar fields on the squashed fuzzy sphere including off-diagonal components,
\begin{equation}
\phi \equiv \Upsilon=\begin{pmatrix}\Upsilon_{+} & \Upsilon_{12}\\
\Upsilon_{21} & \Upsilon_{-}
\end{pmatrix}.
\label{eq:full PSI}
\end{equation}
We denote the scalar fields on $PS^2_N$ with $\Upsilon$ henceforth, to emphasize their stringy nature.
Here $\Upsilon_{\pm}$ correspond to functions on the upper and lower sheet, respectively,
while $\Upsilon_{12}, \Upsilon_{21}$ are naturally interpreted as strings 
connecting these sheets\footnote{Equivalently, one may consider $\Upsilon$ as $\mmu(2)$-valued 
noncommutative gauge field on a single sheet.}.
Then the covariant derivatives acting on the 
string-like modes is
\begin{align*}
D_{i}\Upsilon_{12} =
-i\bar{\theta}_{ii'}^{-1}\left[X^{i'},\Upsilon_{12}\right] 
& \sim\partial_{i}\Upsilon_{12} -i \left(A_{i}^{+}-A_{i}^{-}\right)\Upsilon_{12}\\
D_{i}\Upsilon_{21} =
-i\bar{\theta}_{ii'}^{-1}\left[X^{i'},\Upsilon_{21}\right] 
& \sim\partial_{i}\Upsilon_{21} + i \left(A_{i}^{+}-A_{i}^{-}\right)\Upsilon_{21}
\end{align*}
with 
\begin{eqnarray*}
\vec{A}^{+}-\vec{A}^{-} 
 & = & \kbar^{-1}\begin{pmatrix}K^{1}\\
K^{2}
\end{pmatrix} .
\label{gauge-difference}
\end{eqnarray*}
We note that the off-diagonal
string-like modes couple to the difference $\vec{A}^{+}-\vec{A}^{-}$
of the gauge fields on the two sheets, and behave like charged objects moving in a background
with field strength $F^+ - F^-$.
In particular, the  Laplacian (\ref{eq:BOX-sqashed}) acting on these fields becomes  
\begin{equation}
\Box_{S}\Upsilon_{12} = 
\delta^{ij}D_{i}D_{j}\Upsilon_{12}
\label{eq:gauge-Laplacian}
\end{equation}
in the semi-classical limit.
This is precisely the Hamiltonian for a charged particle moving in a magnetic field,
as studied in section \ref{sec:The-Landau-problem}.
We therefore expect that in the pole limit i.e. near the origin $\vec x = 0$ 
for $r \to \infty$, 
its spectrum should reproduce that of the Landau problem. This will be elaborated below.


\paragraph{Pole limit $\vec{x}=\vec{0}$.}

Near the pole we can expand the field strength in a Taylor series in $x^i$. 
Neglecting terms suppressed by $\mathcal{O}((\frac{x^{i}}{r})^{2})$, we obtain
\begin{align*}
F_{12}^{+}  =0, \qquad 
F_{12}^{-}  =\frac{2}{\kbar}
\end{align*}
and the gauge fields obtained from (\ref{gauge-difference}) are 
\begin{align*}
\vec{A}^{+}-\vec{A}^{-} & =\frac{1}{\kbar}\begin{pmatrix}x^{2}\\
-x^{1}
\end{pmatrix}.
\end{align*}
In particular, $(\vec{A}^{+}-\vec{A}^{-})$ acting
on $\Upsilon_{21}$ corresponds to the field strength 
\begin{equation}
F^+ - F^-=\frac{-2}{\kbar}
\label{eq:Y21-fieldstregth}
\end{equation}
while $(\vec{A}^{-}-\vec{A}^{+})$ acting on $\Upsilon_{12}$
corresponds to the opposite field strength.
Not surprisingly, the two sheets reduce for $r\rightarrow\infty$
to Moyal-Weyl quantum planes $\mathbb{R}_{\theta}^{2}$, with a constant
field $F=\frac{2}{\kbar}$ on the lower sheet. We can of course absorb the field strength in 
any given quantum plane by redefining $\bar\theta^{ij}$, but the difference between the two sheets 
is unambiguous.

\paragraph{Edge limit $(x^{1})^{2}+(x^{2})^{2}=r^2$.}

At the edge, the field strength $F_{ij}$ becomes
\[
F_{12}^{\pm}=\frac{1}{\kbar}
\]
on both sheets, consistent with the fact that the Poisson structures 
on the upper and lower sheet (\ref{eq:theta-edge})  have a smooth transition.

We observe that $F_{12}$ is divergent as $\kbar \to 0$. However this is not a problem,
since we are interested in the physics for fixed $\kbar$ corresponding to 
fixed magnetic field $B$ \eq{B-k-relation}. In other words, we consider the limit 
corresponding to $N\to\infty$ while keeping $B$ or $\kbar$ fixed. 

\subsection{Fuzzy Laplacian and its eigenfunction\label{sub:The-Laplacian-and}}

In the previous chapter we arrived at an interpretation of the matrix Laplacian
$\Box_{S}$ on the squashed fuzzy sphere in the semi-classical limit 
(\ref{eq:gauge-Laplacian}).
Now we return to the fuzzy case, and study this matrix Laplacian exactly.
Comparing it with the Laplacian on the fuzzy sphere (\ref{eq:BOX-sphere}), 
we can write
\[
\Box_{S}=\Box-\frac{1}{\kbar^{2}}\left[X^{3},\left[X^{3},.\right]\right]
\]
In fact we can immediately write down all the eigenvectors and eigenvalues:
they are given by the same fuzzy spherical harmonics $\hat{Y}_{m}^{l}$ which 
diagonalize $\Box$, since $X^{3}$ is proportional to $J^{3}$ and $J^{3}\hat Y_{m}^{l}=m\hat Y_{m}^{l}$.
Thus
\[
\Box_{S}\hat{Y}_{m}^{l}=\frac 1{r^2}\left(l\left(l+1\right)-m^{2}\right)\, \hat{Y}_{m}^{l}.
\]
Note that the spectrum of $\Box_{S}$ is independent of the matrix
dimension $N$, up to the cutoff.
\begin{figure}[h]
\centering{}\includegraphics[scale=0.60]{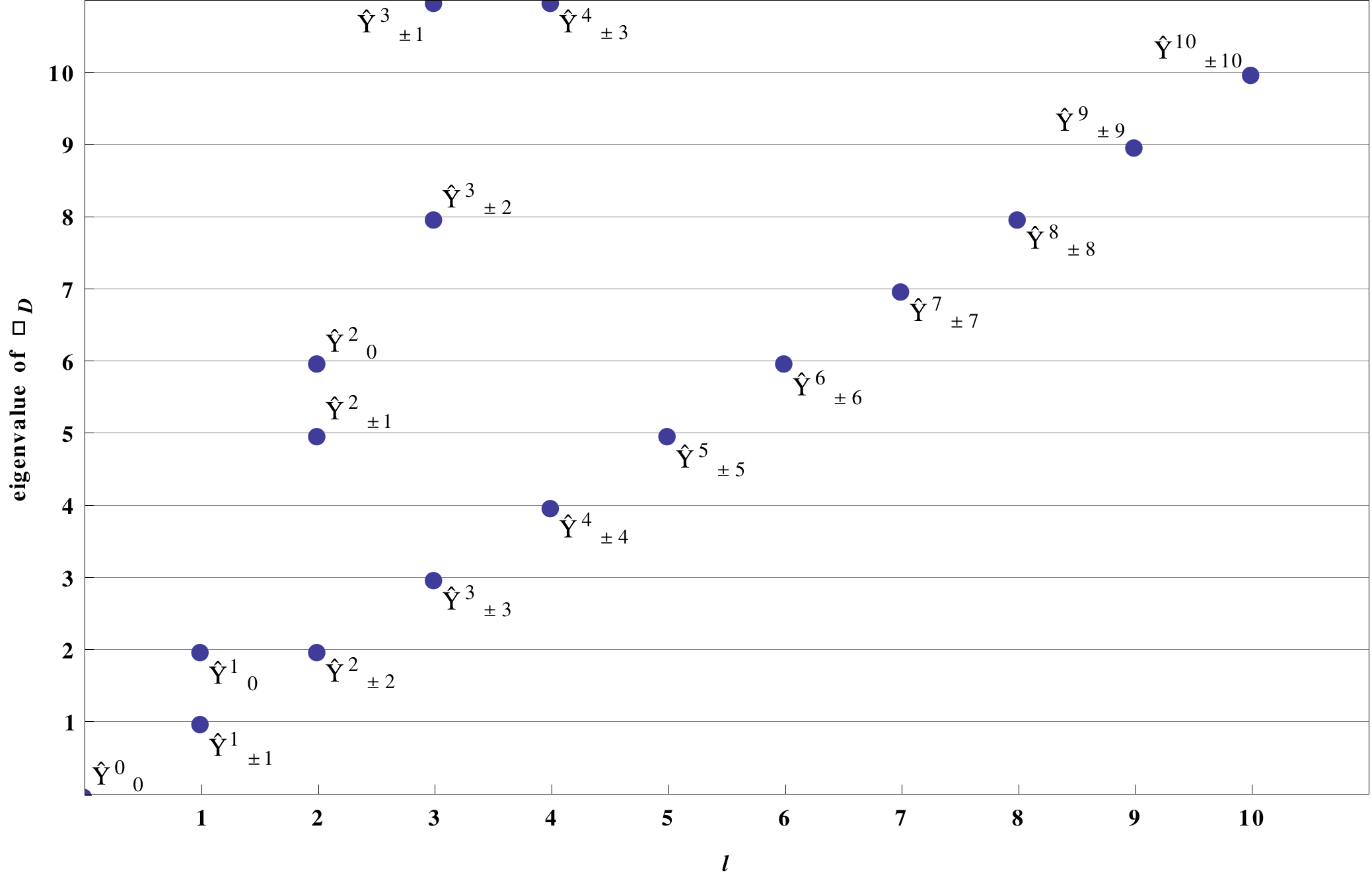}\caption{The first few eigenvalues of $\Box_{S}$ corresponding to $\hat{Y}_{m}^{l}$.
Each eigenvalue where $m\neq0$ is at least twice degenerated, since
$\hat{Y}_{m}^{l}$ and $\hat{Y}_{-m}^{l}$ have the same eigenvalues.
However, this does not mean that the $m=0$ the eigenvalue is not degenerate,
since e.g. $\hat{Y}_{0}^{2}$ and $\hat{Y}_{\pm6}^{6}$
have the same eigenvalue given by 6.\label{fig:Box_D EV}}
\end{figure}

Since the eigenvalues of $\Box_{S}$ are not independent of $m$ anymore
(unlike in $\Box)$, the degeneracy of each eigenvalue has a more
complicated structure than for $\Box$.  Figure \ref{fig:Box_D EV}
shows the lowest  eigenvalues and the corresponding states
$\hat{Y}_{m}^{l}$.

Thus the $\hat{Y}_{m}^{l}$ span the Hilbert space on the squashed fuzzy sphere. However to
identify the string modes $\Upsilon_{12}$, we need to identify those $\hat{Y}_{m}^{l}$,
which have entries exclusively in the upper right block, because of
equation (\ref{eq:full PSI}). In chapter \ref{sub:fuzzy-spherical harmonics}
we saw that with larger $m$, the entries of the $\hat{Y}_{m}^{l}$
are farther away from the main diagonal, thus for $l\simeq l_{max}$
and $\left|m\right|\simeq l$ the $\hat{Y}_{m}^{l}$ have entries
solely in these string domains. Therefore for $m>0$ the $\hat{Y}_{m\simeq l}^{l}$
serve as a basis for the $\Upsilon_{12}$, when $m<0$ for $\Upsilon_{21}$.
These are the string modes we are looking for.

\subsection{Semi-classical limit and string states\label{sub:Semi-classical-limit}}

Having identified the basis for the strings $\Upsilon_{12}$ 
(and $\Upsilon_{21}$) as $\hat{Y}_{m\simeq l}^{l}$ (and $\hat{Y}_{m\simeq-l}^{l}$),
we want to understand their precise relation with the states of the Landau problem, and 
relate the spectrum of $\Box_{S}$ for these string states in the semi-classical limit. 

\begin{figure}
\centering{}\includegraphics[scale=0.52]{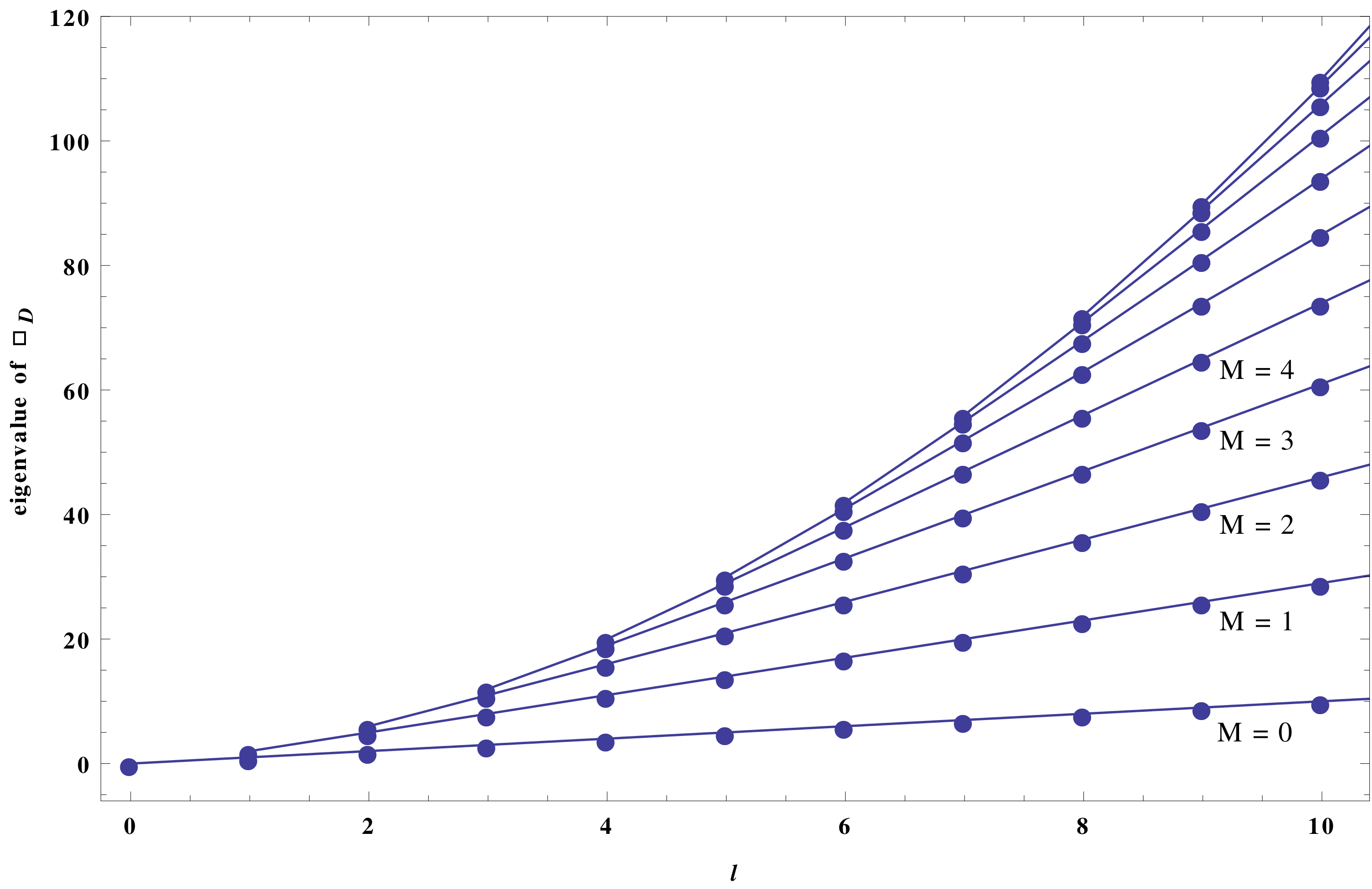}\caption{Levels $M=0,\ldots,10$ for small $l_{max}$, if $N$ is small. If
$N$ is large this domain is where $l$ is small, i.e. these $\hat{Y}_{m}^{l}$
are not suitable as basis for the strings.\label{fig:small N}}
\end{figure}

Recall from chapter \ref{sub:fuzzy-spherical harmonics}
that the quantum numbers for the fuzzy spherical harmonics for fixed matrix dimension $N$
are given by
\begin{eqnarray*}
l & = & 0,1,\ldots,l_{max}, \qquad l_{max}  =  N-1,\\
m & = & l,l-1,\ldots,-l.
\end{eqnarray*}
The distribution of the eigenfunctions in figure \ref{fig:Box_D EV}
already suggests which grouping of the $\hat{Y}_{m}^{l}$ might
be appropriate. The $\hat{Y}_{m}^{l}$ with fixed difference $l-m$
lie on certain lines, as illustrated in figure \ref{fig:small N}.
In order to appropriately describe the $\hat{Y}_{m}^{l}$ states with $l\simeq l_{max}$
and $\left|m\right|\simeq l$ for large $N$, we define two new (small) quantum numbers $L$
and $M$ as the complement of $l$ and $m$ (which are large). 
Let us distinguish the
two cases where $m$ is positive and negative, since they are correlated
to different strings, $\Upsilon_{12}$ and $\Upsilon_{21}$ respectively.
Thus we define
\begin{eqnarray*}
L & := & l_{max}-l\ \in \{0,1,2,...\} \\
M & := & l-m \ \in \{0,1,2,...\} \qquad(\mbox{for}\ m>0),\\
M' & := & l+m \ \in \{0,1,2,...\} \qquad(\mbox{for}\ m<0).
\end{eqnarray*}
Since $\Box_{S}$ in (\ref{eq:gauge-Laplacian}) has the form of a
squared momentum operator $(\partial+A)^{2}$, we multiply it with a factor
of $\frac{1}{2\mu}$ in order to relate $\Box_{S}$ with the Hamiltonian (\ref{eq:Hamiltonian}), where
$\mu$ is the mass of the particle.
Then the eigenvalue equation for $\Box_{S}$ becomes
\[
\frac{1}{2\mu}\Box_{S}\hat{Y}_{m}^{l}=\frac{(l(l+1)-m^{2})}{2\mu r^{2}}\hat{Y}_{m}^{l}.
\]
We can now rewrite this in terms of our new quantum numbers $L$ and
$M$, and get
\begin{eqnarray*}
l(l+1)-m^{2} & \overset{{\scriptstyle m>0}}{=} & \left((N-L-1)(N-L)-(N-1-L-M)^{2}\right)\\
 & = & \left(-1-L-2M-2LM-M^{2}+(1+2M)N\right)\\
 & \overset{{\scriptscriptstyle {\scriptstyle -1-L-2M-2LM-M^{2}\ll N}}}{=} & 2N\left(M+\frac{1}{2}\right)+\mathcal{O}(1)
\end{eqnarray*}
for $m>0$, and 
\[
l(l+1)-m^{2}\quad\overset{{\scriptscriptstyle {\scriptstyle m<0}}}{=}\quad2N\left(M'+\frac{1}{2}\right)+\mathcal{O}(1),
\]
for $m<0$. Here we assume that $N$ is very large while $M,L$ are small, 
as appropriate for the flat (pole) limit.
Accordingly, we define a new basis of string modes as follows:
\begin{align}
 \Upsilon_{(12)}^{L,M} &= \hat{Y}^{l_{max}-L}_{l_{max}-L-M} , \qquad\qquad  l_{max}-L-M >\frac N2 \nn\\
 \Upsilon_{(21)}^{L,M'} &= \hat{Y}^{l_{max}-L}_{-(l_{max}-L) +M'}  , \qquad  -l_{max}+L+M' < -\frac N2
\end{align}
Thus
\begin{align}
\frac{1}{2\mu}\Box_{S}\Upsilon_{(12)}^{L,M} & =\frac{N}{\mu r^{2}}\left(M+\frac{1}{2}\right)\Upsilon_{(12)}^{L,M}
\qquad\mbox{for} \ L=0,1,2,\ldots\label{eq:1/2mu Box_D EV}\\
\frac{1}{2\mu}\Box_{S} \Upsilon_{(21)}^{L,M'}  & =\frac{N}{\mu r^{2}}\left(M'+\frac{1}{2}\right) \Upsilon_{(21)}^{L,M'} 
\qquad\mbox{for} \ L=0,1,2,\ldots.\nonumber 
\end{align}
We can  compare this to  the eigenvalue equation (\ref{eq:E_perp n_r}) of the
Landau problem in chapter \ref{sec:The-Landau-problem} 
\begin{align*}
H\left|\chi_{n_{r},n_{l}}\right\rangle  & =\hbar\omega_{c}\left(n_{r}+\frac{1}{2}\right)\left|\chi_{n_{r},n_{l}}\right\rangle 
   \qquad\text{for \ensuremath{q<0}}\qquad\text{with }n_{l}=0,1,2,\ldots\\
H\left|\chi_{n_{r},n_{l}}\right\rangle  & =\hbar\omega_{c}\left(n_{l}+\frac{1}{2}\right)\left|\chi_{n_{r},n_{l}}\right\rangle 
   \qquad\text{for \ensuremath{q>0}}\qquad\text{with }n_{r}=0,1,2,\ldots
\end{align*}
Note that we have both charged sectors $q=\pm 1$ realized 
at the same time, by the $\Upsilon_{(12)}$ and $\Upsilon_{(21)}$ respectively.
Therefore we can identify 
\begin{align*}
M & \equiv n_{r}\\
M' & \equiv n_{l}
\end{align*}
 and  
\begin{eqnarray*}
\frac{N}{\mu r^{2}} & = & \hbar\omega_{c}.
\label{cyclotron-match}
\end{eqnarray*}
Recall that we are using Planck units $\hbar,c=1$, and
the coupling constant is set to  $q=\pm1$. Thus using the
definition of $\omega_{c}$ from (\ref{eq:omega_c}) and  $\kbar\simeq\frac{2r^{2}}{N}$ from 
(\ref{kbar-def}) for large $N$, we obtain
\begin{align}
\frac{2}{\kbar} & =B
\label{B-k-relation}
\end{align}
in the semi-classical limit. 
This is indeed precisely the field strength acting
on strings $\Upsilon_{12}$ connecting the upper to the lower sheet
near the poles as we have seen in (\ref{eq:Y21-fieldstregth}).
Therefore we have found complete agreement between the Landau problem and the 
string states on the squashed fuzzy sphere in the planar limit.

\begin{figure}
\centering{}\includegraphics[scale=0.45]{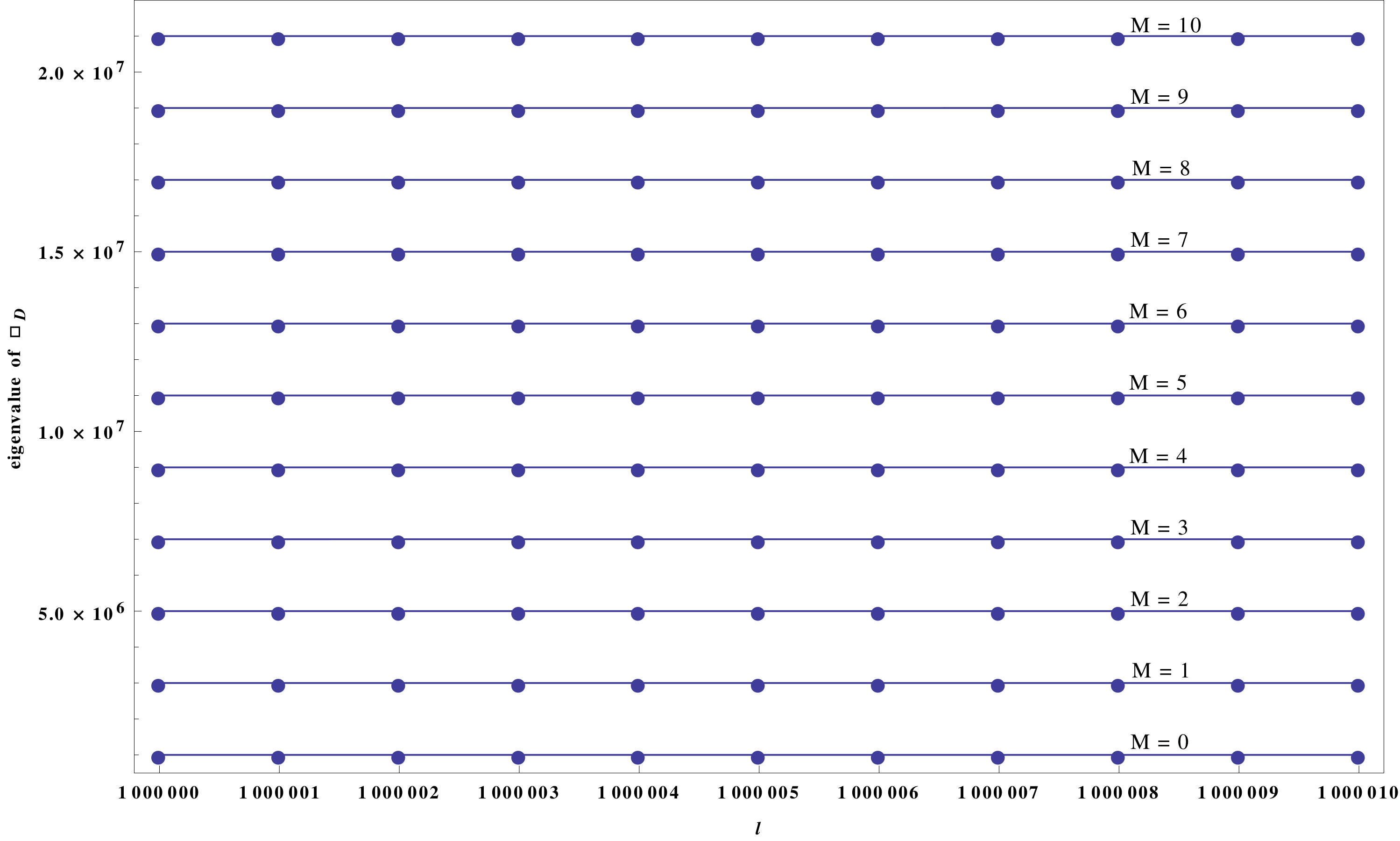}\caption{Levels $M=0,\ldots,10$ for large $l$, if $N$ is large. For large
$N$ these levels are approximately constant over this interval of
$l$. Notice that the lowest level $M=0$ is not 0, but has an
offset, compatible with the result in equation (\ref{eq:1/2mu Box_D EV}).
\label{fig:large N}}
\end{figure}

To illustrate this, we display in figure \ref{fig:small N} and \ref{fig:large N}
the eigenfunctions of $\frac{1}{2\mu}\Box_{S}$
for small and large $N$, and match these with the eigenstates of the Landau problem.
The lowest Landau level is given by the wave functions
$\chi_{0,n}$, where $n$ is either $n_{l}$ or $n_{r}$ depending
on the sign of the $\vec{B}$-field. The lowest level of $\frac{1}{2\mu}\Box_{S}$
is given by the $\hat{Y}_{l}^{l}$ or $\hat{Y}_{-l}^{l}$ for very large
$l$. Thus the $\chi_{n_{r},0}$ should be identified with $\hat{Y}_{-l}^{l}$,
and the $\chi_{0,n_{l}}$ with $\hat{Y}_{l}^{l}$. 
Note that these are the highest and lowest weight states in the algebra of functions $Mat(N,\C)$.
In particular, $\chi_{n_{r}=0,n_{l}=0}$
can be identified with both $\hat{Y}_{l_{max}}^{l_{max}}= \Upsilon_{(12)}^{0,0}$ 
or $\hat{Y}_{-l_{max}}^{l_{max}}=  \Upsilon_{(21)}^{0,0}$.
The reason for this doubling is that we have both charged sectors $q=\pm 1$ realized 
at the same time, by the $\Upsilon_{12}$ and $\Upsilon_{21}$ respectively.
 We can thus 
identify the states in the various Landau levels as
\begin{align}
 \chi_{0,n_{l}} &\ \longleftrightarrow\ \hat{Y}_{l_{max}-n_{l}}^{l_{max}-n_{l}} = \Upsilon_{(12)}^{n_l,0}, \nn\\
 \chi_{1,n_{l}} &\ \longleftrightarrow\ \hat{Y}_{l_{max}-n_{l}-1}^{l_{max}-n_{l}} = \Upsilon_{(12)}^{n_l,1}, \nn\\
 \chi_{2,n_{l}} &\ \longleftrightarrow\ \hat{Y}_{l_{max}-n_{l}-2}^{l_{max}-n_{l}} = \Upsilon_{(12)}^{n_l,2}
 \label{states-identific-1}
\end{align}
and so forth. 
Similarly for the opposite charges,
\begin{align}
 \chi_{n_{r},0}&\ \longleftrightarrow\ \hat{Y}_{-(l_{max}-n_{r})}^{l_{max}-n_{r}} = \Upsilon_{(21)}^{n_r,0}, \nn\\
 \chi_{n_{r},1}&\ \longleftrightarrow\ \hat{Y}_{-(l_{max}-n_{r})+1}^{l_{max}-n_{r}} = \Upsilon_{(21)}^{n_r,1}
  \label{states-identific-2}
\end{align}
and so forth. 
The  quantum number labeling the degenerate states
in a Landau level can be identified using the operator  
$J_z^{(\rm ad)} = \kappa^{-1}[X_3,.]$,
which corresponds to angular momentum around the $z$ axis on the fuzzy sphere.
We can compute its eigenvalue either directly from the $SU(2)$ quantum numbers
\begin{align}
 J_z^{(\rm ad)} \Upsilon_{(12)}^{n_l,0} &= (N -1-n_l) , \qquad
  J_z^{(\rm ad)} \Upsilon_{(21)}^{n_r,0} = -(N-1-n_r)
\end{align}
or ín the semi-classical limit $N\to\infty$ as follows
\begin{align}
 J_z^{(\rm ad)} \Upsilon_{(12)}^{n_l,0}  &= \kappa^{-1}\big(X_3 \Upsilon_{(12)} - \Upsilon_{(12)} X_3\big) \nn\\
  &= \sqrt{C_N}\Big(1-\frac 1{2r^2}((X^1)^2 + (X^2)^2) + O\big(\frac{x^4}{r^4}\big)\Big) \Upsilon_{(12)} \nn\\
 &\quad -  \sqrt{C_N}\Upsilon_{(12)}\Big(-1 + \frac 1{2r^2}((X^1)^2 + (X^2)^2) + O\big(\frac{x^4}{r^4}\big)  \Big) \nn\\
  & \sim \frac N2\Big(2 - \frac 1{r^2}((x^1)^2 + (x^2)^2)+ O(\frac{x^4}{r^4}) \Big) \Upsilon_{(12)} .
\end{align}
Neglecting the $O(\frac{x^4}{r^4})$ terms and
taking into account the above identifications,  we find 
\begin{align}
 ((x^1)^2 + (x^2)^2) \, \Upsilon_{(12)}^{n_l,0}  = \frac {r^2}{\sqrt{C_N}} (n_l+1) \, \Upsilon_{(12)}^{n_l,0}
 = \frac 2B (n_l+1) \, \Upsilon_{(12)}^{n_l,0} \ .
\end{align}
Hence these states are localized on circles around the origin with radius measured by $n_l$,
just like the states $\chi_{0,n_{l}}$ in the Landau problem \eq{landau-wavefunct}.
The analogous statement for $\chi_{n_{r},0}$ completes the identification 
of the harmonics on the squashed fuzzy sphere with those of the Landau problem.

\begin{figure}
\centering{}\includegraphics[scale=0.15]{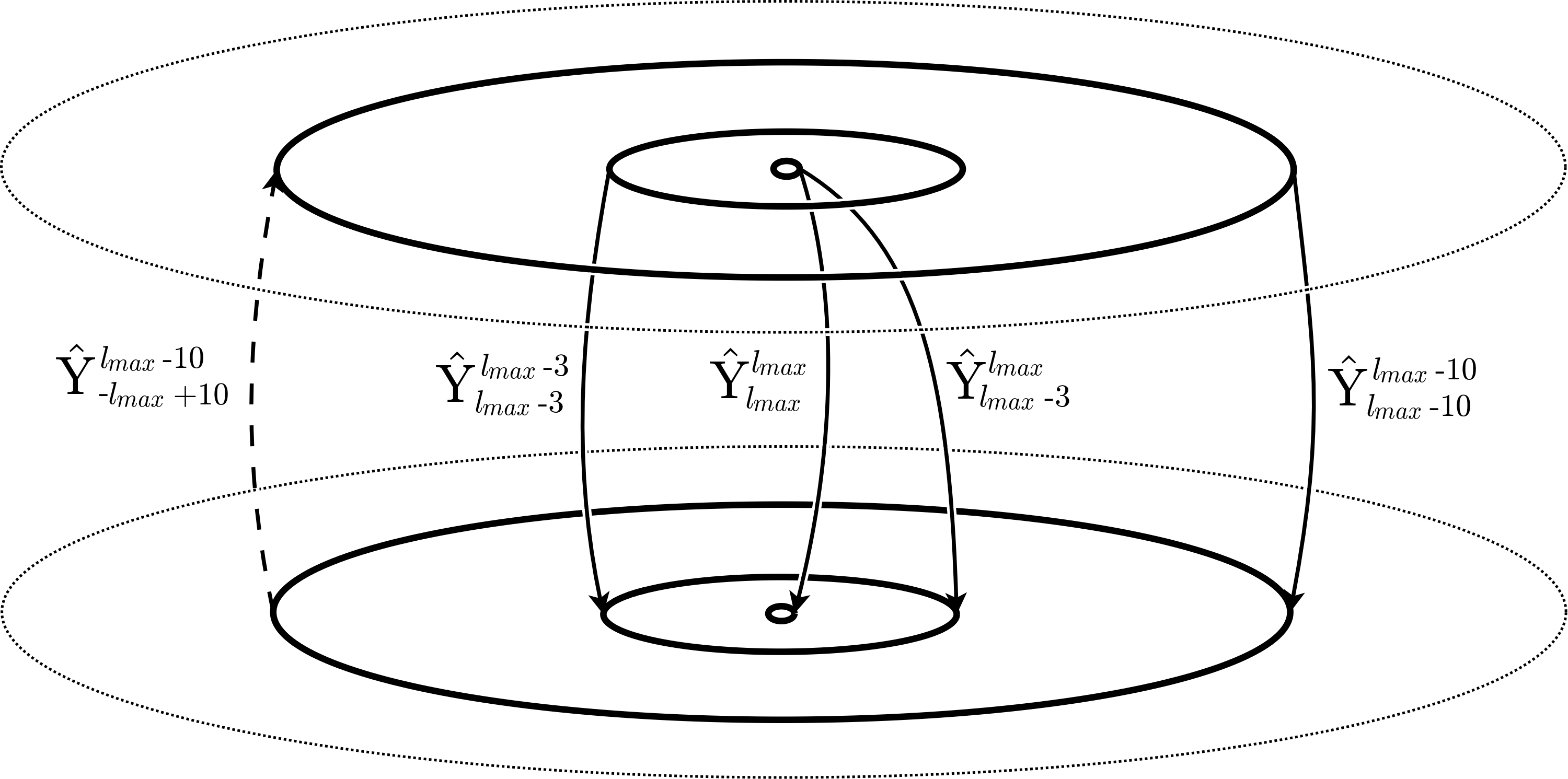}\caption{The strings $\Upsilon_{12}$ and $\Upsilon_{21}$ should be thought
of as connecting the upper and lower sheet, creating transitions from
one to the other. The solid arrows create transitions from the upper
to the lower sheet and should be identified with $\Upsilon_{12}$,
while the dashed with $\Upsilon_{21}$. Strings creating transitions
from pole to pole will have very large quantum number
 $|m| \approx l_{max}$ in terms of $\hat{Y}_{\pm m}^{l_{max}}$.
}
\label{fig:strings}
\end{figure}

Finally we can exhibit the stringy interpretation of these matrix states as links between the sheets.
The states $\hat{Y}_{l_{max}}^{l_{max}}= \Upsilon_{(12)}^{0,0}$ 
or $\hat{Y}_{-l_{max}}^{l_{max}}=  \Upsilon_{(21)}^{0,0}$ can be written 
explicitly as follows
\begin{align}
 \Upsilon_{(12)}^{0,0} &= \left|\frac{N-1}{2}\right\rangle \left\langle -\frac{N-1}{2}\right|  \nn\\
 \Upsilon_{(21)}^{0,0} &= \left|-\frac{N-1}{2}\right\rangle \left\langle \frac{N-1}{2}\right| .
 \label{coherent-states}
\end{align}
Here the extremal weight states $|\pm \frac{N-1}{2}\rangle$ are the coherent states localized at the 
north and south pole of the fuzzy sphere, hence at the origin of the two fuzzy disks. 
This makes the interpretation of the $\Upsilon_{(12)}^{0,0}$ as strings connecting the two sheets 
at the origin manifest, and vindicates the identification
with $\chi_{0,0}$ \eq{states-identific-1}, \eq{states-identific-2}.
Although the expressions of the other states in terms of 
coherent states is more complicated, it is clear 
that they can be thought of as slightly extended strings localized at circles around the origin.
This is illustrated in figure \ref{fig:strings}.

\subsection{Dirac operator}

For completeness, we briefly discuss also the  Dirac operator on the squashed fuzzy sphere.
$\slashed{D}$ is naturally defined by 
\[
\slashed{D}=\frac{1}{\kbar}\left(\sigma_{1}\otimes\left[X^{1},.\right]+\sigma_{2}\otimes\left[X^{2},.\right]\right)
\]
with $X^{1},X^{2}$ from (\ref{eq:full X}). We can compute its square
\begin{eqnarray}
\slashed{D}^{2} & = & \frac{1}{\kbar^{2}}\sigma_{i}\sigma_{j}\otimes\left[X^{i},\left[X^{j},\cdot\right]\right]\\
 & = & \Box_{S}\otimes\mathbbm{1}_{2}-\frac{1}{\kbar^{2}}\left[X^{3},\cdot\right]\otimes\sigma_{3} \nn\\
 & = & \Box \otimes\mathbbm{1}_{2}-\frac{1}{\kbar^{2}}\big([X^{3},\cdot] + \frac 12\sigma_{3}\big)^2 
  + \frac{1}{4\kbar^{2}}
  \label{Dsquare}
\end{eqnarray}
using the commutation relations of the fuzzy sphere. 
The last form allows to find immediately the eigenvalues and eigenfunctions 
following \cite{Steinacker:2014lma}:
Decomposing the space of functions 
$Mat(N,\C)  = \oplus_{l=0}^{N-1} \C^{2l+1}$ with basis $|l,m_l\rangle$
and passing to the 
total angular momentum basis of $\C^2\otimes \C^{2l+1}$ labeled by $j,l,m_j$, the eigenvalues are
\begin{align}
 E_{jlm_j}^2 &= 4 l(l+1) - 4 m_j^2 + 1  = 4(l+\frac 12-m_j)(l+\frac 12+m_j)
\end{align}
where $m_j = m+s$, and $s$ is the eigenvalue of $\frac 12 \s_3$.
Hence for each $l\in \{0,1,2,...,  N-1\}$, there is pair of zero modes with extremal 
weights $m_j=\pm(l+\frac 12)$,  which can be written as 
\begin{align} 
 \Psi_+ = |\uparrow\rangle |l,l\rangle , \qquad \Psi_- = |\downarrow\rangle |l,-l \rangle .
\end{align} 
Thus the fermionic zero modes correspond to the
extremal weight states in the angular momentum decomposition of $Mat(N,\C)$. 
In particular for $L=0$ or $l=l_{max}$, these can again be written in terms of 
coherent states as in (\ref{coherent-states}). More generally,
these zero modes can  be interpreted as fermionic strings, 
linking the two opposite sheets at or near the origin.

On the other hand, the second form in \eq{Dsquare} 
allows to easily take the semi-classical (pole) limit as in the previous section. 
Using the analogous procedure as for the Laplacian before, we obtain 
\[
\frac 1{2\mu}\slashed{D}^{2} \Upsilon_{(12)}^{L,M} 
=\hbar\omega_{c}\left(M+\frac{1}{2}-\frac{\sigma_{3}}{2}\right)\Upsilon_{(12)}^{L,M}
\]
in the large $N$ limit using  $\frac{2N}{r^{2}} = \hbar\omega_{c}$ for (\ref{cyclotron-match}), 
and similarly for the $\Upsilon_{(21)}^{L,M'}$.
Thus for these string states,
$\frac 1{2\mu}\slashed{D}^{2}$ reproduces the Hamiltonian of the Landau levels 
including spin (\ref{eq:landauwithspin}), for $g=1$.
In particular, we can understand the above fermionic zero modes as fermions in the lowest Landau level
with appropriate orientation of the spin.
Remarkably, they are exact zero modes even for finite $N$. 
For generalizations we refer the reader to \cite{Steinacker:2014lma}.

\section{Conclusion}

We have identified string-like modes among the fuzzy spherical harmonics, which connect 
the upper with the lower hemisphere. On the squashed fuzzy sphere,
these behave like charged objects moving under the influence of a magnetic field.
In the large $N$ limit, this field becomes approximately constant
in the vicinity of the origin resp. the north and south poles, and 
the lowest string-like  modes behave like charged point-like objects. 
In particular, we have identified the lowest Landau levels among these fuzzy spherical harmonics,
providing an organization of the space of functions in terms of string-like modes.

Our results illustrate the well-known fact that noncommutative field theory is much richer than 
ordinary gauge theory, and behaves more like a string theory rather than a field theory.
The present example provides a particularly
clear identification of such string modes in a non-trivial background, 
in a simple  finite-dimensional setting. 
It illustrates how non-trivial backgrounds in matrix models can be understood 
quantitatively in the semi-classical limit. The present example is related to the 
new solutions of (deformed) $\cN=4$ SYM and the IKKT matrix model 
\cite{Steinacker:2014lma,Steinacker:2014eua}, which 
could be analyzed in a similar way. 
More generally,
a systematic use of analogous string-like modes in the study of noncommutative field theory
might help to illuminate various issues and problems in this context.

\paragraph{Acknowledgments.}

The work of H.S. is supported by the Austrian Science Fund (FWF) grant P24713.

\bibliographystyle{diss}
\bibliography{bibl.bib}

\end{document}